# Bond relaxation and electronic properties of T-type WTe$_2$/MoS$_2$ heterostructure using BOLS BB and BC model

Hongrong Qiu, Hanze Li, Jiannan Wang, Yunhu Zhu, Maolin Bo*

Key Laboratory of Extraordinary Bond Engineering and Advanced Materials Technology (EBEAM) of Chongqing, Yangtze Normal University, Chongqing 408100, China

Corresponding Author: *E-mail: bmlwd@yznu.edu.cn (Maolin Bo)

**Abstract**

We combined the bond-order-length-strength (BOLS) binding energy(BB) and bond charge (BC) models and the topological concept to obtain the nonbonding, bonding, and antibonding states of the T-type WTe$_2$/MoS$_2$ heterostructure. We found that the electronic probability and electronic dispersion in the valence band of the WTe$_2$/MoS$_2$ heterostructure can be determined precisely based on electronic entropy. The energy-band projection method and electronic entropy are remarkable approaches for analyzing the electronic properties of various structures based on DFT calculations. This study provides a new way to describe the electronic properties of T-type heterostructures and calculate the electron and bonding state probabilities.

# 1. Introduction

Two-dimensional transition metal sulfides (TMDs) have attracted considerable attention due to their advantageous properties, such as high current-carrying mobility and wide adjustable band gap.[1-3] TMDs are used in many applications, including semiconductor devices and photocatalysis.[4] TMDs have a sandwich-like structure, with a transition metal element layer in the middle and two chalcogen elements on both sides.[5] Different kinds of TMDs materials have different electrical and optical properties,[6] and specific properties can be obtained by forming TMDs heterostructures.[7] By growing $MoSe_2/WSe_2$ heterostructures, excitons can be formed in the $MoS_2/MoSe_2$ heterostructure, and the interlayer interaction of the $MoS_2/WS_2$ heterostructure can be observed. Ullah et al. investigated the growth of $MoSe_2/WSe_2$ lateral heterostructures.[8] Ceballos et al. reported the formation of long-lived indirect excitons in an $MoS_2$-$MoSe_2$ heterostructure.[9] Yelgel et al. showed the $WS_2/MoS_2$ heterostructure has a stable structure and excellent electronic properties.[10] The results of these studies suggest that the formation of a two-dimensional TMDs heterostructure may change the electronic state, and form the excitons, allowing their application in various devices.[11] Most two-dimensional TMDs heterostructures are stacked into different shapes and connected by van der Waals forces between layers.[12] However, there are few studies on vertical heterojunctions connected by chemical bonds. In optoelectronic devices, vertical heterojunctions can better satisfy electron migration and provide a larger working area for optoelectronic devices.[13] Therefore, we studied the T-type heterostructure of $WTe_2/MoS_2$.

In this study, we constructed a T-type $WTe_2/MoS_2$ heterostructure and modeled its structure using density functional theory (DFT) calculations.[14] Then, we investigated bond relaxation and electronic properties of the T-type $WTe_2/MoS_2$ heterostructure using the energy-band projection method. In particularly, we analyzed the density of state (DOS) and band structure in each energy range along each Brillouin path. We used the concept of electronic entropy to express the electronic dispersion in the valence band of the T-type $WTe_2/MoS_2$ heterostructure. This energy-band projection method and the concept of electronic entropy provide a new approach to study the electronic properties of various structures.

## 2. Methods

### 2.1 DFT calculation

We aimed to analyze the energetics, electronic properties and the atomic structure of the T-type WTe$_2$/MoS$_2$ heterostructure. All structural relaxation and electronic properties of the WTe$_2$/MoS$_2$ heterostructure were calculated by CASTEP.[15] We used the HSE06 hybrid density function to describe the electron exchange and correlation potential;[16] the cut-off energy was set to 650 eV, and $3 \times 2 \times 1$ $k$-point grids were used. In the calculations, the energy converged to $10^{-6}$ eV, and the force on each atom converged to <0.01 eV/Å.

### 2.2 BOLS- BB model

The Hamiltonian is given by

$$H = \left[ -\frac{\hbar^2 \nabla^2}{2m} + V_{atom}(r) \right] + V_{cry}(r)(1+\Delta_H)$$

(1)

$V_{atom}(r)$ is the intra-atomic potential of the atom, $V_{cry}(r)$ is the potential of the crystal, and the interaction potential changes with the coordination environment and during chemical reactions. The electronic binding energy (BE) of the $v$th energy band $E_v(0)$ and $E_v(B)$ are

$$E_v(0) = -\langle v,i| -\frac{\hbar^2 \nabla^2}{2m} + V_{atom}(r) |v,i\rangle$$

(2)

$$E_v(x) - E_v(0) = -\alpha_v - \sum_j f(k) \cdot \beta_v = -\alpha_v(1+\sum_j \frac{f(k) \cdot \beta_v}{\alpha_v}) \cong -\alpha_v(1+\Delta_H) \propto \langle E_x \rangle$$

(3)

The $E_v(x)$ and $E_v(0)$ are the energy levels of atoms and an isolated atom, respectively. The $\alpha_v$ and $\beta_v$ contributes to the width of the energy band. In the localized band of core levels, $\beta_v$ is very small, so $\alpha_v$ determines the energy band of the core levels.

$$\alpha_v = -\langle v,i|V_{cry}(r)|v,i\rangle \propto \langle E_b \rangle$$

$$\beta_v = -\langle v,i|V_{cry}(r)|v,j\rangle; \sum_j f(k)\beta_v \propto \langle E_x - E_b \rangle$$

(4)

The $|v,i\rangle$ represents the wave function. The periodic factor $f(k)$ is the form of $e^{ikr}$, while $k$ is the wave vector. The $\beta$ is dependent on the overlap between orbitals centered at two neighboring atoms.

The bond energy $E_x$ uniquely determines the core-level BE shift:[17]

$$z_x = \frac{12}{\left\{8\ln\left(\frac{2\Delta E_v'(x) - \Delta E_v(B)}{\Delta E_v(B)}\right) + 1\right\}} \quad (\Delta E_v(x) \geq 0)$$

(5)

where $z_x$ is the atomic coordination number of an atom in the $x$th atomic layer from the surface.

$$\frac{\Delta E_v(x)}{\Delta E_v(B)} = \frac{\Delta E_v'(x) + \Delta E_v(B)}{\Delta E_v(B)}$$

(6)

$$\frac{\Delta E_v(w_x)}{\Delta E_v(w_B)} \propto \frac{E_x}{E_b} = \gamma = 1 + \Delta_H = c_x^{-m} = \left(\frac{d_x}{d_b}\right)^{-m}$$

(7)

$$\frac{\Delta E_v(w_x)}{\Delta E_v(w_B)} \cong \frac{\Delta E_v'(x) + \Delta E_v(B)}{\Delta E_v(B)} = \gamma$$

(8)

$$\Delta V_{cry}(r) = V_{cry}(r)(1 + \Delta_H) = \gamma V_{cry}(r) = (Z+1)\frac{1}{4\pi\varepsilon_0}\sum_i \langle v,i|\frac{e^2}{r_i}|v,i\rangle$$

(9)

$$E_v(x) - E_v(0) \cong -\langle v,i|V_{cry}(r)(1 + \Delta_H)|v,i\rangle = -(Z+1)\frac{1}{4\pi\varepsilon_0}\sum_i \langle v,i|\frac{e^2}{r_i}|v,i\rangle$$

(10)

$Z$ is the initial atom charge (neutral $Z = -1$ (isolated atom), positively $Z = 0$ (bulk atoms), positively $Z = +|\delta\gamma|(\gamma > 0)$ (charged positive atoms) and negatively $Z = -|\delta\gamma|(\gamma < 0)$ (charged negative atoms). Thus the core-electron BE shifts will be 0,

$-\sum_i \langle v,i|\frac{(1+|\delta\gamma|)}{4\pi\varepsilon_0}\frac{e^2}{r_i}|v,i\rangle, -\sum_i \langle v,i|\frac{1}{4\pi\varepsilon_0}\frac{e^2}{r_i}|v,i\rangle, -\sum_i \langle v,i|\frac{(1-|\delta\gamma|)}{4\pi\varepsilon_0}\frac{e^2}{r_i}|v,i\rangle$ and initially

neutral of isolated atom, bulk atoms, singly charged positive and negative atoms, respectively.[18] **Eq. 7** provides estimates for the bond energy $E_x$, bond length $d_x$ and $\Delta E_v(w_B) \cong \Delta E_v(B)$ is the spectral full width of the bulk component ($w_B$) of the $v$th energy level; The width of the BE shift for the surface component ($w_x$) of the $v$th energy level is $\Delta E_v(w_x) = \Delta E_v(w_B) + \Delta E_v'(x)$; actual spectral intensities and shapes, however, are subject to polarization effects and measurement artifacts. We can calculate the chemisorption and defect-induced interface bond energy ratio $\gamma$ with the known reference value of $\Delta E_v'(x) = E_v(x) - E_v(B)$, $\Delta E_v(B) = E_v(B) - E_v(0)$ and $\Delta E_v(x) = E_v(x) - E_v(0)$ derived from the surface via DFT calculations and XPS analysis. Hence, we obtain

$$\delta\gamma = \frac{\Delta E'_v(x)}{\Delta E_v(w_B)} = \frac{\Delta E'_v(x) + \Delta E_v(B)}{\Delta E_v(B)} - 1 = \gamma - 1 \quad \text{(RBER)}$$

(11)

$$\delta\varepsilon_x = d_x / d_b - 1 = \gamma^{-1} - 1 \quad \text{(RLBS)}$$

(12)

$$\delta E_d = (E_i / d_i^3) / (E_b / d_b^3) - 1 = \gamma^4 - 1 \quad \text{(RBED)}$$

(13)

$$\Delta E'_v(x) \cong -\delta\gamma \langle v,i|V_{cry}(r)|v,i\rangle \cong -\sum_j f(k)\langle v,i|V_{cry}(r)|v,j\rangle \propto \langle E_x - E_b \rangle$$

(14)

Thus, one can drive the interface binding energy ratio (RBER) parameter $\delta\gamma$. If $\delta\gamma < 0$, the binding energy is reduced, the potential of the crystal and the bond is weakened. Conversely, if $\delta\gamma > 0$, the binding energy increases, the potential of the crystal and the bond becomes stronger. The relative local bond strain (RLBS) $\delta\varepsilon_x$ indicates the relative contraction of the atomic bond length $d_x$. The relative bond energy density (RBED) $\delta E_d$ is the energy density of the atomic bond with energy $E_i$.

## 3. Results and discussion

### 3.1 Electronic properties of T-type WTe$_2$/MoS$_2$ heterostructure

We obtained the electronic and structural properties of the heterostructure via DFT calculations. As is shown in **Fig. 1**, the structure of the T-type WTe$_2$/MoS$_2$ heterostructure. It can be seen from the figure that the structure is T-shaped. The lattice parameters of the T-type WTe$_2$/MoS$_2$ heterostructure are presented in **Table 1** (a = 6.235 Å, b = 12.980 Å, c = 14.638 Å), and the optimized atomic coordinate results are presented in **Table 2**. The band structure of the WTe$_2$/MoS$_2$ heterostructure is shown in **Fig. 2**. The band gap of the T-type WTe$_2$/MoS$_2$ is 0.328 eV. This structure has an indirect band gap. In addition, we observed a flat band in the band gap of the WTe$_2$/MoS$_2$ heterostructure. In **Fig. 2**, a very distinct horizontal band with almost no dispersion appears when the Fermi level ($E_f = 0$) is near zero. It can be concluded that the flat band can adjust the electron states of the band structure.

The total DOS of the T-type $WTe_2/MoS_2$ heterostructure is shown in the **Fig. 3**. It can be seen from the figure that the zero domain appears Fermi level ($E_f = 0$). The **Fig. 3** also presents the electronic distribution of the valence band maximum (VBM) of the T-type $WTe_2/MoS_2$ heterostructure has an energy of 0 eV and is mainly distributed on the Fermi level. The electron in the conduction band minimum (CBM) is at an energy of approximately 0.015 eV. The PDOS are shown in **Fig. 4**. From the PDOS, we can see that there is a peak at −0.236 eV in the electron density distribution near the Fermi level in the valence band, and a peak at 0.673 eV in the conduction band. We calculate the atomic local DOS of the $WTe_2/MoS_2$ heterojunction of Mo ($4p4d5s$), S ($3s3p$), W ($6s5d$) and Te ($5s5p$) atoms. As shown in **Fig. 5**, the $WTe_2/MoS_2$ heterostructure of main electron contributions to the conduction band come from the Mo $4d$ and S $3p$ orbitals. Moreover, the W and Te atoms have less electron contribution to the Fermi level.

### 3.2 Energy-band projection method

We used a new method to analyze the energy distribution near the Fermi surface of the T-type $WTe_2/MoS_2$ heterostructure. Also, we can use this method to analyze the probability of electrons appearing in various energy ranges along each Brillouin path. In the following article, I will use electronic entropy to explain the reason. First, we normalized the DOS,[19] extracted the points with energy between −3 eV and 3 eV in the DOS, and listed the extracted energy as the basis; at the same time, we extracted the energy points and equally divided the corresponding points into 3000 points; after the equalization, the point sequence was integrated and normalized. Second, we stratified the normalized DOS. The point list was integrated piecewise, sorted with a unit of 0.2 eV, and divided into 11 levels (−3 eV ~ 3 eV). Different levels are represented by different colors for probability of electrons. Then, we conducted hierarchical processing on the band structure. The points with energy between −3 eV and 3 eV, and the corresponding data were extracted completely and the energy range obtained in the second step was matched with the color. The different between Energy-band projection method and the Fatband software for 2D band projection, is that the Fatband software using the *K*-point projection;[20] while Energy-band projection method take the energy projection for subsequent analyses.

**Fig. 6** is displayed in a flow picture of the projection, DOS is projected on the band structure diagram based on energy, and different colors are used to represent the electron probabilities. According to Shannon, for a discrete probability distribution, the entropy is defined as follows: [21]

$$H(X) = -\sum_{x_i \in X} P(x_i) \log P(x_i).$$

(15)

For the entropy of electrons, the random variable $X$ represents the energy range corresponding to the point on the path of the Brillouin region, $x_i$ represents one of the energy ranges, $P(x_i)$ represents the probability of the electron appearing in a certain energy range. The electronic entropy corresponding to each energy range can be obtained using the data displayed in **Table 3**. Near the valence band (−3 eV ~ 0 eV), the electron entropy was 0.116 bit. The entropy near the shallow energy level of the valence band (−0.6 eV ~ 0 eV) was calculated to be 0.018 bit. And the conduction band (0 eV ~ 3 eV), the electron entropy was 0.090 bit. The entropy near the shallow energy level of the conduction band (0 eV ~ 1.2 eV) is 0.013 bit.

The order of the electronic entropy values in each range was consistent with the order of the corresponding electron appearance probability. In **Fig. 7**, the energy band structure of the processed DOS projection. Therefore, we can see that the color is closer to black (0 bit), the smaller the entropy value, the smaller the probability of electrons appearing, and the color is closer to red (0.055 bit), the greater the entropy value, and the greater the probability of electrons appearing. It can be clearly seen that around the Fermi level, especially the energy range of the valence band is −0.6 eV to 0 eV, the probability of the occurrence of electrons is generally low. This indicates that the shallow energy level of the valence band is very close to the cavitation-bound state of the valence band in the range of −0.6 eV to 0 eV. The electrons or holes in the shallow energy level of the valence band are ionized at a slightly higher temperature (e.g., room temperature) to become free electrons in the conduction band and free holes in the valence band, which promote the conduction of electricity.

### 3.3 Bonding states of T-type WTe$_2$/MoS$_2$ heterostructure

We considered a free quantum particle of charge $e$. We confined the charge $e$ in one dimension and subjected it to periodic boundary conditions, i.e., a particle on a ring. Thus, the free Hamiltonian of the system takes the following forms:[22]

$$\begin{cases} H = -\frac{\hbar^2}{2m}\left(\nabla - \frac{q}{\hbar c}A(\vec{r})i\right)^2 \\ \quad = \frac{1}{2}\left(-i\partial_\phi - A(\vec{r})\right)^2 \quad (e=1, h=1, c=1), \\ \psi_n(\phi) = \frac{1}{\sqrt{2\pi}}\exp(in\phi), \end{cases}$$

(16)

where $\hbar$ is Planck's constant, $q$ is the amount of charge, $c$ is the speed of light, $\vec{r}$ is the radius of the electron, and $\phi$ is the field.

To understand a particle on a ring, we express the field $\phi$ as mapping

$$\phi: S_1 \to S_1, \quad \tau \to \phi(\tau),$$

from the unit circle $S_1$ into another circle. Mappings of this type can be assigned to a winding number $W$ (see **Fig. 8**). The number of times $\phi(\tau)$ winds around the unit circle as $\tau$ progresses from 0 to $\beta$ : $\phi(\beta) - \phi(0) = 2\pi W$. Here, we noted that the $A(\vec{r})$-dependent term in the action,

$$S_{top}[\phi] \equiv iA(\vec{r})\int_0^\beta d\tau\dot{\phi} = iA(\vec{r})\big(\phi(\beta) - \phi(0)\big) = i2\pi W A(\vec{r}).$$

(17)

Only the index of topological sector $\phi$. $S_{top}$ is the topological sector of the field contribution.

The atomic bond relaxation is expressed based on the BOLS-BB model as follows:[23]

$$E_i \propto V_{cry}(\vec{r}_{ij}) = qA(\vec{r}_{ij}) = qS_{top}[\phi]/i2\pi W$$

$$\Delta E'_v(x) \cong -\sum_j f(k)\langle v,i|qA(\vec{r}_{ij})|v,j\rangle = -\sum_j f(k)\langle v,i|qS_{top}[\phi]/i2\pi W|v,j\rangle \propto \Delta E_i \tag{18}$$

$$\frac{E_i}{E_B} = \left(\frac{1/d_i}{1/d_B}\right)^m \propto \frac{V_{cry}(r_{ij})}{V_{cry}(r_B)} = \gamma; \begin{cases} \gamma > 1, deepening\ the\ potential\ well \\ \gamma < 1, strengthening\ the\ potential\ Energy\ Barrier \end{cases} \tag{19}$$

Then,

$$V_{cry}(r_{ij}) = \frac{1}{4\pi\varepsilon_0}\iiint d^3 r_{ij} \iiint d^3 r_{ij} \frac{\Delta\rho_{hole}(r_{ij})\Delta\rho_{electron}(r_{ij})}{r_{ij}}, \tag{20}$$

where $E_i$ is the single bond energy, $\Delta\rho(r_{ij})$ is the deformation charge density, $d_i$ is the bond length of the atom, $V_{cry}(r_{ij})$ is the crystal potential, $B$ represents bulk atoms, and $m$ is the bond nature indicator. In **Eq. 18 and 19**, chemical bonds are associated with the topological effects and quantum fields. In **Eq. 20**, the potential function $\gamma V_{cry}(r_{ij})$ may become deeper ($\gamma > 1$) or shallower ($\gamma < 1$) than the corresponding $V_{cry}(r_B)$ of the specific constituent. **Eq. 21** describes the relationship between the deformation charge density $\Delta\rho(r_{ij})$ and the crystal potential $V_{cry}(r_{ij})$. This method has been applied to a variety of structures such as two-dimensional Sb/MoSe$_2$ and BN/SiC vdW heterostructures.[23, 24]

**Fig. 9** was depicted the deformation charge density. We believe that the formation of the WTe$_2$/MoS$_2$ heterostructure is mainly due to the contributions of bonds, electrons, and charges.[25] Using **Eq. 19**, we calculated crystal potentials of the antibonding states, nonbonding states and bonding states of WTe$_2$/MoS$_2$ heterostructure.[26] **Table 4** presents the electronic states of the three chemical bonds calculated by DFT calculations. The nonbonding states, bonding states, and antibonding states were displayed in the deformation charge density. The dark-red area indicates the bonding state. The light-red area is a nonbonding state. The white area represents the antibonding state. Furthermore, we analyzing the deformation charge density to definite the atomic bonding information.

## 4. Conclusion

We used DFT to calculate the constructed T-typed $WTe_2/MoS_2$ heterostructure and combined the BOLS-BC model and the topological concept to obtain the nonbonding state, bonding state, and antibonding state of the T-typed $WTe_2/MoS_2$ heterostructure. Furthermore, the energy band projection method was used to study the electronic properties of the $WTe_2/MoS_2$ heterostructure. Using the formula of electronic entropy, the valence band (−3 eV ~ 0 eV); the shallow energy level of the valence band (−0.6 eV ~ 0 eV); the conduction band (0 eV ~ 3 eV) and the shallow energy level of the conduction band (0 eV ~ 1.2 eV) were calculated to represent the electronic probability of the $WTe_2/MoS_2$ heterostructure, respectively. The energy-band projection method and the concept of electronic entropy provide a new approach for studying the electronic properties of various structures.

**Acknowledgment**

The Scientific and Technological Research Program of Chongqing Municipal Education Commission (KJQN201901424), the Chongqing Natural Science Foundation project (cstc2020jcyjmsxmX0524)

**Figure and Table Captions**

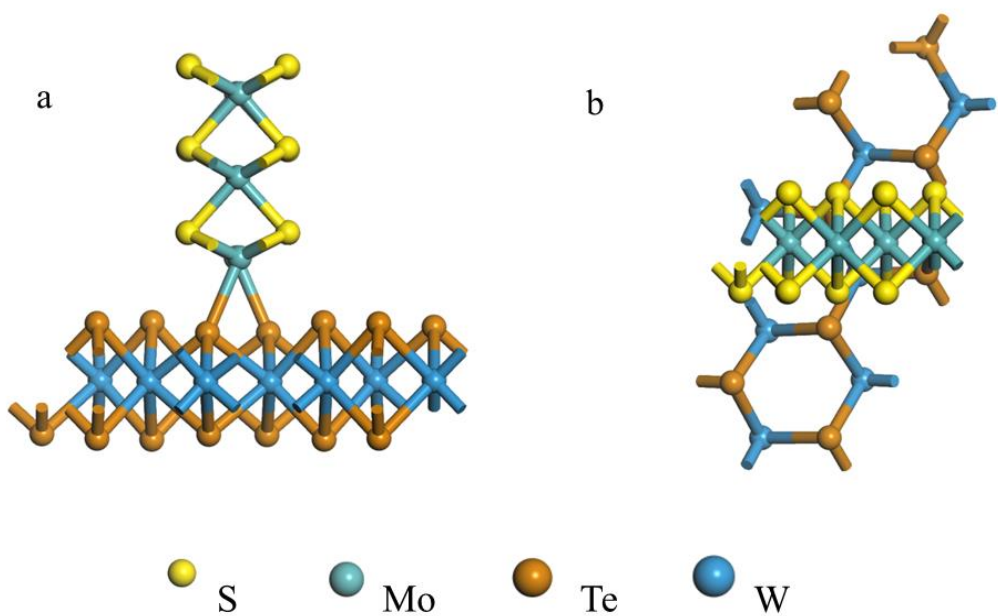

- S  ● Mo  ● Te  ● W

**Fig. 1** Front view (a) and top view (b) of T-type $WTe_2/MoS_2$ heterostructure. Yellow, green, orange, and blue balls represent sulfur, molybdenum, tellurium, and tungsten atoms, respectively.

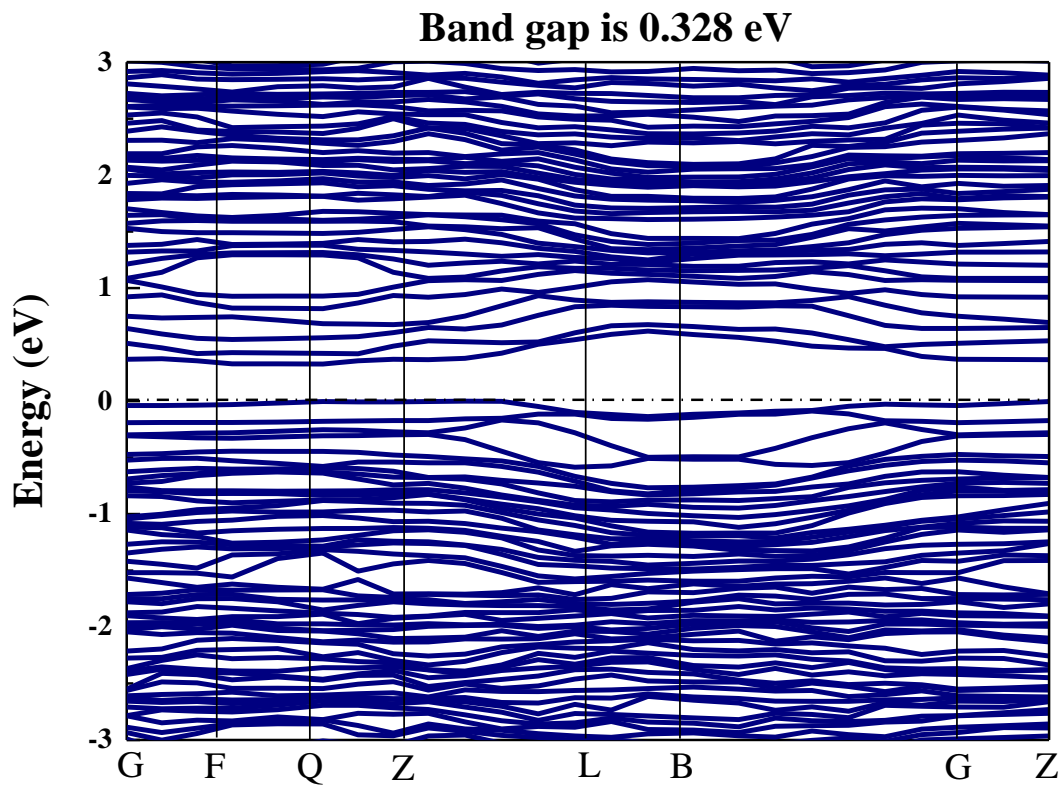

**Fig. 2** Band structure of T-type $WTe_2/MoS_2$ heterostructure

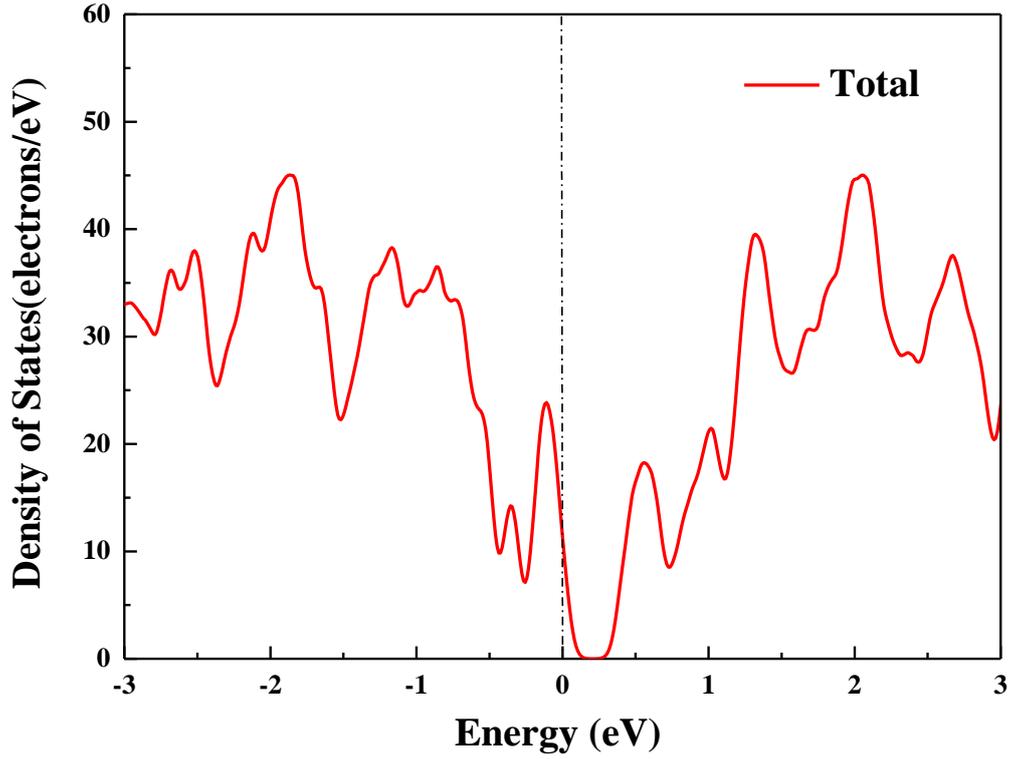

**Fig. 3** Total density of states of T-type $WTe_2/MoS_2$ heterostructure

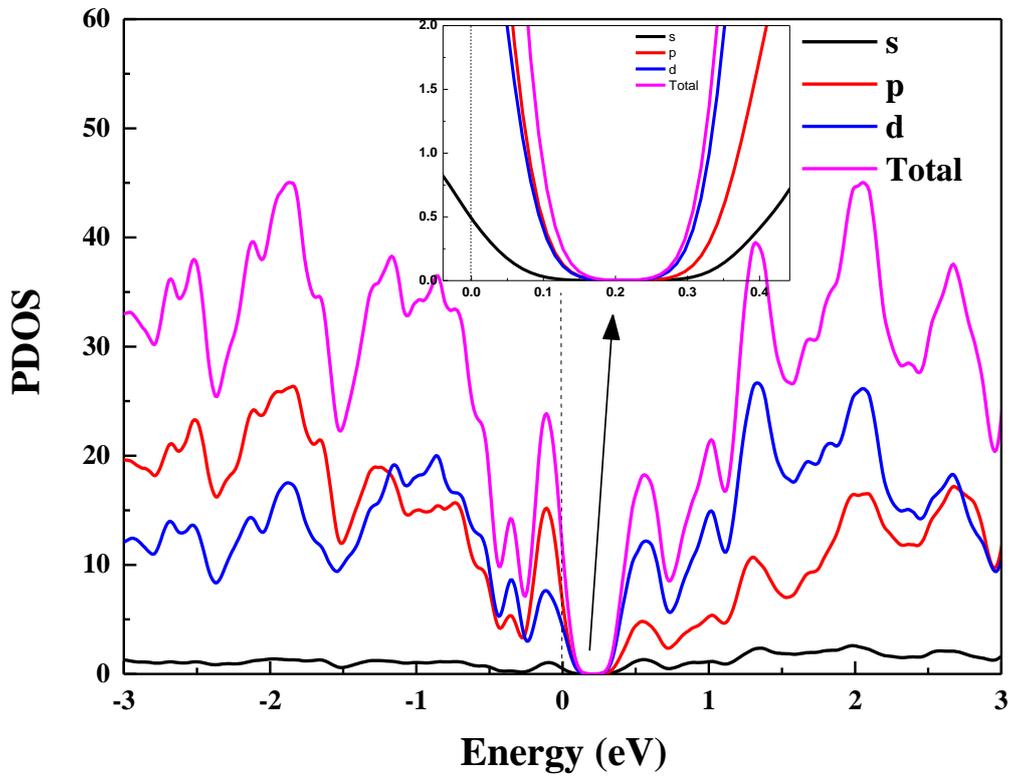

**Fig. 4** Partial density of states (PDOS) of T-type $WTe_2/MoS_2$ heterostructure

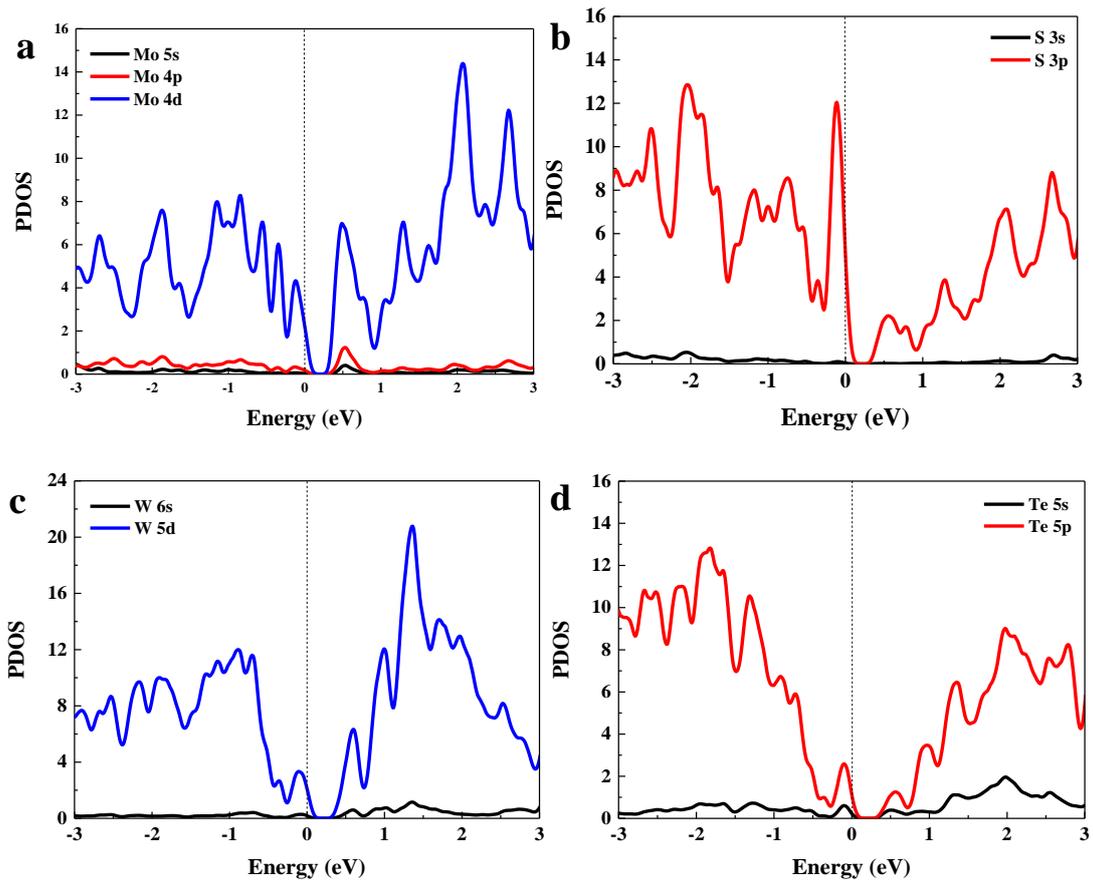

**Fig. 5** Partial density of states (PDOS) of Mo, S, W and Te atoms

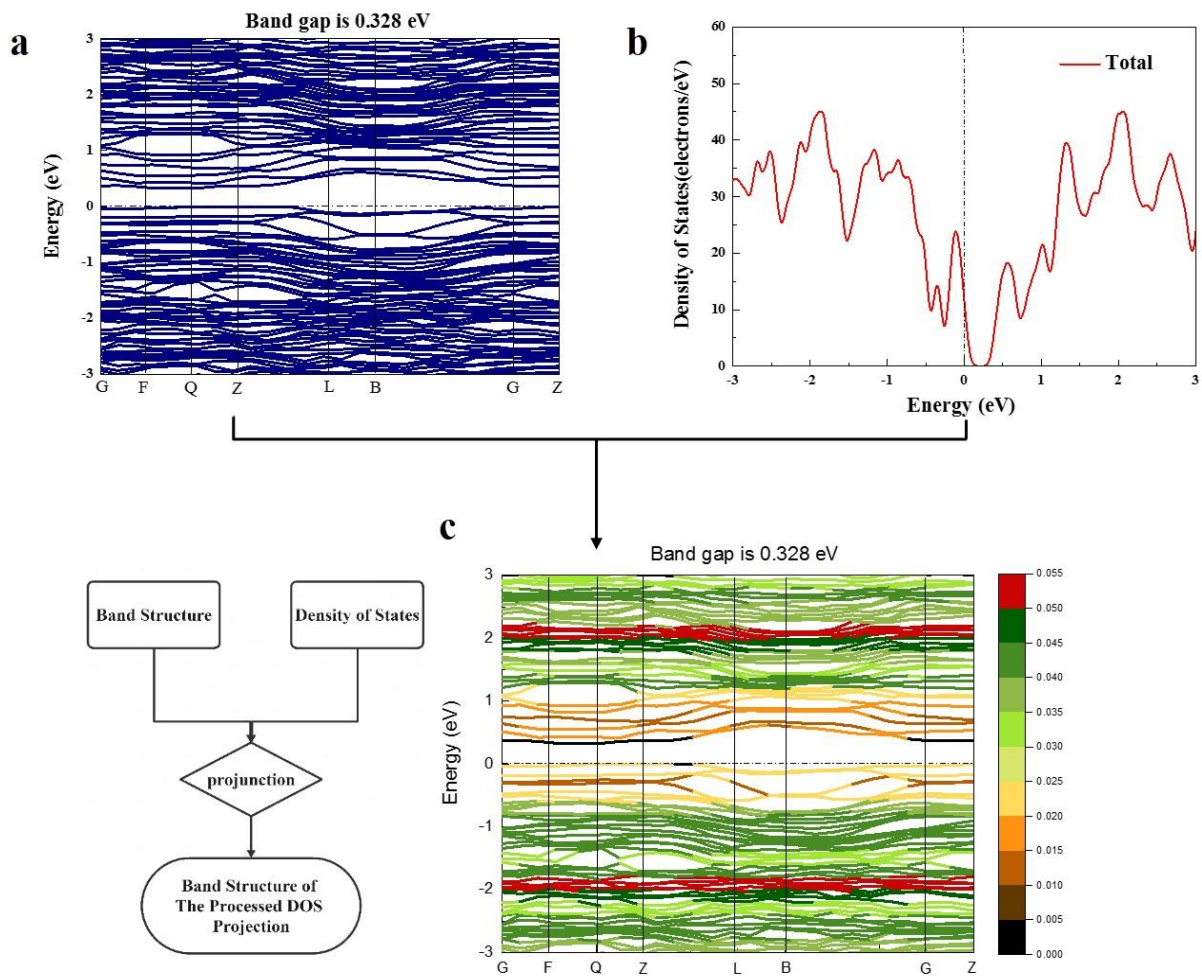

**Fig. 6** Projection flow chart; (a) band structure, (b) total density of states, and (c) band structure of the processed DOS projection.

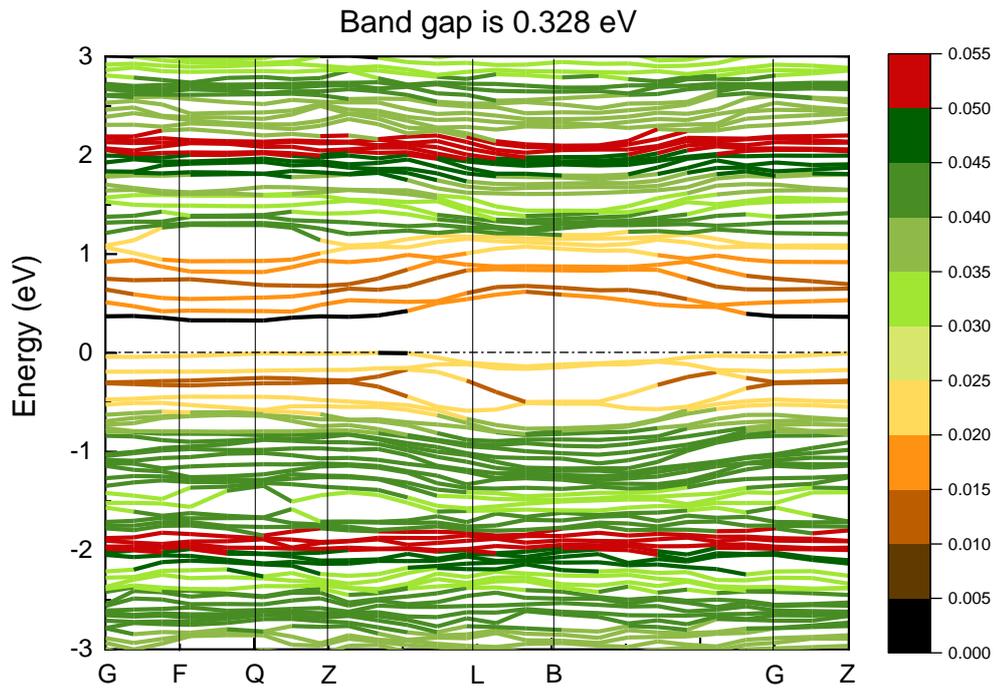

**Fig. 7** Band structure of the processed DOS projection. Colors closer to black and red represent a lower and higher probability, respectively, that the electrons appear in the corresponding energy range on the path of the Brillouin zone.

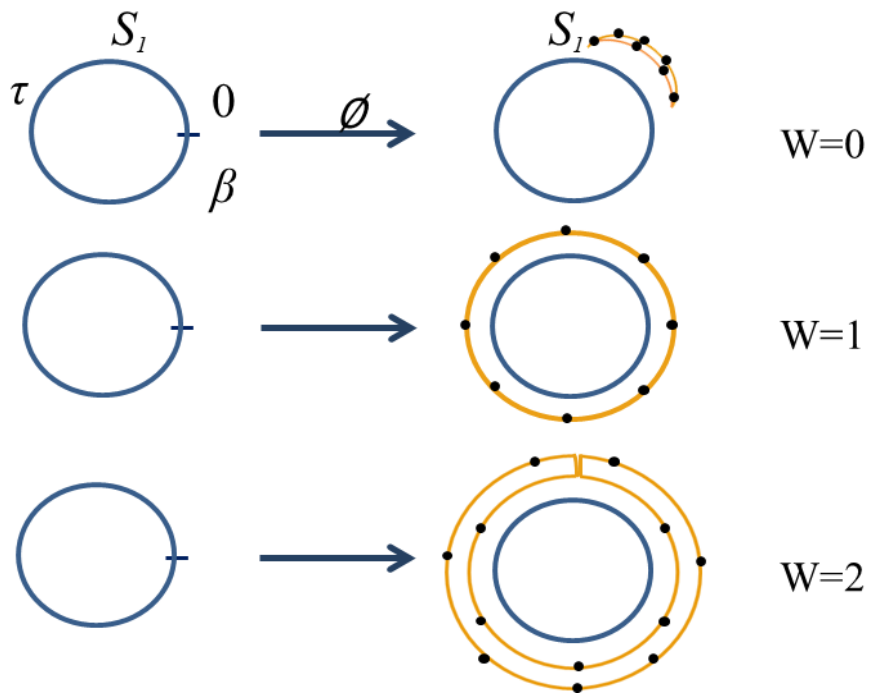

**Fig. 8** Mappings $\phi: S_1 \to S_1$ of different winding numbers (W).

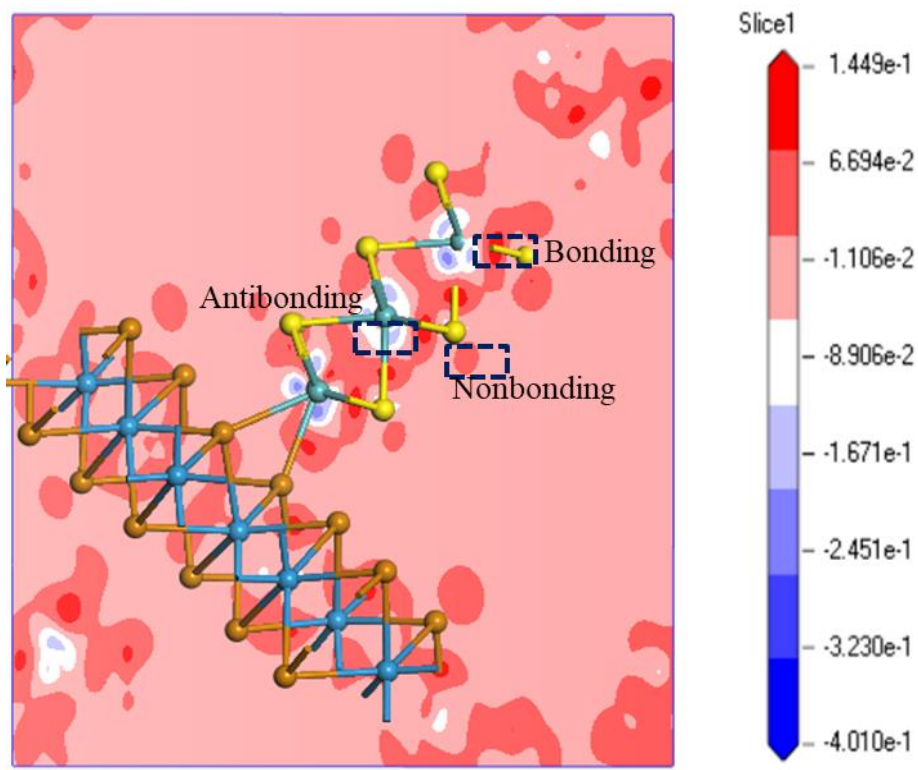

**Fig. 9** Deformation charge density of WTe$_2$/MoS$_2$ heterostructure.

**Table 1** Lattice parameters of T-type $WTe_2/MoS_2$ heterostructure.

| Structure | Angle | | | Lattice parameter (Length) | | |
|---|---|---|---|---|---|---|
| | α (deg) | β (deg) | γ (deg) | a (Å) | b (Å) | c (Å) |
| $WTe_2/MoS_2$ | 90.00 | 90.00 | 76.10 | 6.235 | 12.980 | 14.638 |

**Table 2** T-type $WTe_2/MoS_2$ heterostructure atomic fractional coordination after optimization.

| Atom | x | y | z |
|---|---|---|---|
| Te1 | 0.687 | 0.000 | 0.111 |
| Te2 | 0.114 | 0.140 | 0.101 |
| Te3 | 0.548 | 0.280 | 0.099 |
| Te4 | 0.972 | 0.426 | 0.109 |
| Te5 | 0.398 | 0.574 | 0.110 |
| Te6 | 0.827 | 0.719 | 0.099 |
| Te7 | 0.254 | 0.859 | 0.101 |
| W8 | 0.338 | 0.000 | 0.229 |
| W9 | 0.786 | 0.148 | 0.230 |
| W10 | 0.232 | 0.274 | 0.227 |
| W11 | 0.628 | 0.418 | 0.226 |
| W12 | 0.044 | 0.582 | 0.226 |
| W13 | 0.505 | 0.726 | 0.227 |
| W14 | 0.935 | 0.851 | 0.230 |
| Te15 | 0.686 | 0.000 | 0.347 |
| Te16 | 0.116 | 0.140 | 0.355 |
| Te17 | 0.541 | 0.281 | 0.357 |
| Te18 | 0.974 | 0.422 | 0.334 |
| Te19 | 0.396 | 0.577 | 0.334 |
| Te20 | 0.823 | 0.718 | 0.358 |
| Te21 | 0.257 | 0.859 | 0.355 |
| Mo22 | -0.025 | 0.504 | 0.501 |
| Mo23 | 0.479 | 0.496 | 0.501 |
| S24 | 0.284 | 0.373 | 0.556 |
| S25 | 0.795 | 0.372 | 0.555 |
| S26 | 0.168 | 0.627 | 0.556 |
| S27 | 0.657 | 0.627 | 0.556 |
| Mo28 | 0.228 | 0.500 | 0.678 |
| Mo29 | 0.728 | 0.499 | 0.678 |
| S30 | 0.041 | 0.375 | 0.742 |
| S31 | 0.541 | 0.373 | 0.739 |
| S32 | 0.418 | 0.624 | 0.743 |
| S33 | 0.914 | 0.626 | 0.739 |
| Mo34 | -0.022 | 0.503 | 0.863 |
| Mo35 | 0.481 | 0.495 | 0.863 |
| S36 | 0.287 | 0.375 | 0.924 |
| S37 | 0.789 | 0.372 | 0.921 |
| S38 | 0.162 | 0.626 | 0.922 |
| S39 | 0.663 | 0.623 | 0.925 |

**Table 3** Probability of the energy range and corresponding color.

| Probability(Integral value) | Energy range | Color |
|---|---|---|
| 0.038 | (−3.0 eV ~ −2.8 eV) | 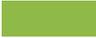 |
| 0.040 | (−2.8 eV ~ −2.6 eV) | 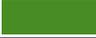 |
| 0.041 | (−2.6 eV ~ −2.4 eV) | 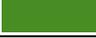 |
| 0.035 | (−2.4 eV ~ −2.2 eV) | 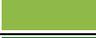 |
| 0.046 | (−2.2 eV ~ −2.0 eV) | 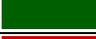 |
| 0.053 | (−2.0 eV ~ −1.8 eV) | 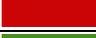 |
| 0.042 | (−1.8 eV ~ −1.6 eV) | 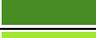 |
| 0.030 | (−1.6 eV ~ −1.4 eV) | 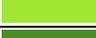 |
| 0.041 | (−1.4 eV ~ −1.2 eV) | 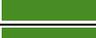 |
| 0.042 | (−1.2 eV ~ −1.0 eV) | 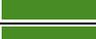 |
| 0.042 | (−1.0 eV ~ −0.8 eV) | 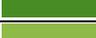 |
| 0.037 | (−0.8 eV ~ −0.6 eV) | 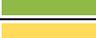 |
| 0.020 | (−0.6 eV ~ −0.4 eV) | 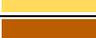 |
| 0.013 | (−0.4 eV ~ −0.2 eV) | 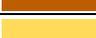 |
| 0.023 | (−0.2 eV ~ 0.0 eV) | 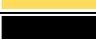 |
| 0.003 | (0.0 eV ~ 0.2 eV) | 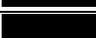 |
| 0.001 | (0.2 eV ~ 0.4 eV) | 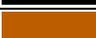 |
| 0.014 | (0.4 eV ~ 0.6 eV) | 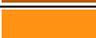 |
| 0.018 | (0.6 eV ~ 0.8 eV) | 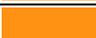 |
| 0.019 | (0.8 eV ~ 1.0 eV) | 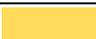 |
| 0.023 | (1.0 eV ~ 1.2 eV) | 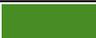 |
| 0.043 | (1.2 eV ~ 1.4 eV) | 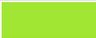 |
| 0.034 | (1.4 eV ~ 1.6 eV) | 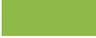 |
| 0.037 | (1.6 eV ~ 1.8 eV) | 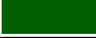 |
| 0.046 | (1.8 eV ~ 2.0 eV) | 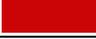 |
| 0.050 | (2.0 eV ~ 2.2 eV) | 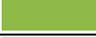 |
| 0.035 | (2.2 eV ~ 2.4 eV) | 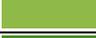 |
| 0.037 | (2.4 eV ~ 2.6 eV) | 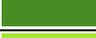 |
| 0.042 | (2.6 eV ~ 2.8 eV) | 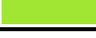 |
| 0.030 | (2.8 eV ~ 3.0 eV) | 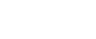 |

**Table 4** Chemical bonding states, deformation charge density $\Delta\rho_i(r_{ij})$, and potential functions $V_{cry}(r_{ij})$ of the van der Waals heterojunctions obtained using the calculated bond charge ($\varepsilon_0 = 8.85 \times 10^{-12} C^2 N^{-1} m^{-2}, e = 1.60 \times 10^{-19} C$)

|  | Mo/S ($r_{ij}$=d/2=2.407/2 Å) |
|---|---|
| $\Delta\rho^{hole}(r_{ij})(e/Å^3)$ | $-4.010 \times 10^{-1}$ |
| $\Delta\rho^{bonding-electron}(r_{ij})(e/Å^3)$ | $1.449 \times 10^{-1}$ |
| $\Delta\rho^{nonbonding-electron}(r_{ij})(e/Å^3)$ | $6.694 \times 10^{-2}$ |
| $\Delta\rho^{antibonding-electron}(r_{ij})(e/Å^3)$ | $-8.906 \times 10^{-2}$ |
| $V_{cry}^{nonbonding}(r_{ij})(eV)$ | $-0.976$ |
| $V_{cry}^{bonding}(r_{ij})(eV)$ | $-2.112$ |
| $V_{cry}^{Antibonding}(r_{ij})(eV)$ | $1.298$ |